\documentclass[runningheads]{llncs}
\usepackage[T1]{fontenc}
\usepackage{graphicx}
\usepackage{multirow}
\usepackage{cite}
\usepackage{subcaption}
\usepackage{enumitem}
\usepackage{booktabs}
\usepackage{hyperref}
\usepackage{lscape}
\usepackage{dingbat}
\usepackage{color, colortbl}
\usepackage{algorithm}
\usepackage{algpseudocode}
\usepackage{amssymb}
\usepackage{comment}
\usepackage{float}
\usepackage{bbm}
\usepackage{amsmath}\usepackage{xcolor}
\definecolor{tmi_blue}{cmyk}{100,0.37,0.0,0.15}
\definecolor{tmi_red}{cmyk}{0,1,1,0}
\usepackage{siunitx}
\usepackage{threeparttable}
\makeatletter
\renewcommand{\section}{\@startsection{section}{1}{\z@}%
  {-1.5ex plus -0.5ex minus -0.2ex}
  {0.8ex plus 0.4ex}
  {\normalfont\Large\bfseries}}

\renewcommand{\subsection}{\@startsection{subsection}{2}{\z@}%
  {-1.2ex plus -0.4ex minus -0.2ex}
  {0.6ex plus 0.3ex}
  {\normalfont\large\bfseries}}

\renewcommand{\subsubsection}{\@startsection{subsubsection}{3}{\z@}%
  {-1.0ex plus -0.3ex minus -0.2ex}
  {0.4ex plus 0.3ex}
  {\normalfont\normalsize\itshape}}
\makeatother

\setlist[enumerate]{itemsep=0.2em, topsep=0.2em, parsep=0em, partopsep=0em}

\definecolor{tmi_blue}{cmyk}{100,0.37,0.0,0.15}
\usepackage{hyperref}
\hypersetup{
    colorlinks=true,
    linkcolor=tmi_blue,
    citecolor=tmi_blue,
    urlcolor=tmi_blue
}
\begin{document}

\title{EMedNeXt: An Enhanced Brain Tumor Segmentation Framework for Sub-Saharan Africa using MedNeXt V2 with Deep Supervision}
\titlerunning{EMedNeXt: An Enhanced Brain Tumor Segmentation Framework for SSA}

%

\author{Ahmed Jaheen\inst{1, 2}\thanks{Equal contribution.}
\and Abdelrahman Elsayed\inst{2}$^{\star}$
\and Damir Kim\inst{2}$^{\star}$
\and\\Daniil Tikhonov\inst{2}
\and Matheus Scatolin\inst{2}
\and Mohor Banerjee\inst{2}
\and Qiankun Ji\inst{2}
\and Mostafa Salem\inst{2}
\and Hu Wang\inst{2}
\and Sarim Hashmi\inst{2}
\and Mohammad Yaqub\inst{2}}

\authorrunning{A. Jaheen et al.}

\institute{
The American University in Cairo (AUC), Cairo, Egypt\\
\and Mohamed bin Zayed University of Artificial Intelligence, Abu Dhabi, UAE\\
\email{{firstname.lastname}@mbzuai.ac.ae}\\
\url{https://mbzuai.ac.ae}
}
%
\maketitle             

\begin{abstract}
Brain tumors, particularly gliomas, pose a significant global health burden, with magnetic resonance imaging (MRI) serving as the primary tool for diagnosis and disease monitoring. However, the current standard for tumor quantification through manual segmentation of multi-parametric MRI is time-consuming, requires expert radiologists, and is often infeasible in under-resourced healthcare systems. This problem is especially pronounced in low-income regions, where MRI scanners are of lower quality and radiology expertise is scarce, leading to incorrect segmentation and quantification. In addition, the number of acquired MRI scans in Africa is typically small. To address these challenges, the BraTS-Lighthouse 2025 Challenge focuses on robust tumor segmentation in sub-Saharan Africa (SSA), where resource constraints and image quality degradation introduce significant shifts. In this study, we present \textbf{\textit{EMedNeXt}}—an enhanced brain tumor segmentation framework based on MedNeXt V2 with deep supervision and optimized post-processing pipelines tailored for SSA. \textbf{\textit{EMedNeXt}} introduces three key contributions: a larger region of interest, an improved nnU-Net v2-based architectural skeleton, and a robust model ensembling system. Evaluated on the hidden validation set, our solution achieved an average LesionWise DSC of \textbf{0.897} with an average LesionWise NSD of \textbf{0.541} and \textbf{0.84} at a tolerance of 0.5mm and 1.0mm, respectively. Our GitHub repository can be accessed here: \href{https://github.com/BioMedIA-MBZUAI/EMedNeXt-BraTS-SSA-2025}{Project Repository}.



\end{abstract}

\section{Introduction}

Gliomas are the most aggressive and prevalent type of primary brain tumor, characterized by poor survival rates and high morbidity. This is particularly severe in pediatric cases, where only about 20\% of patients survive beyond two years after diagnosis \cite{gliomas}. MRI is central in detecting and monitoring gliomas, providing high-resolution 3D insights into brain tissue and tumor subregions. Accurate segmentation of these tumor regions from multi-modal MRI scans is critical for determining treatment options, assessing response to therapy, and guiding long-term follow-up \cite{owrangi2018mri}.

However, manual segmentation remains the clinical standard, which is time-consuming, resource-intensive, and susceptible to human variability. These challenges are significantly magnified in low-resource settings, such as sub-Saharan Africa, where limited access to radiologists and reliance on lower-quality MRI machines can result in poor diagnostic outcomes. Furthermore, publicly available brain MRI datasets from African populations are scarce, which limits the generalizability of current state-of-the-art machine learning models \cite{bratsafrica}.

To address these limitations, the BraTS-Lighthouse 2025 Challenge introduced a new task focused on brain tumor segmentation in sub-Saharan African (SSA) patients hosted by the Medical Image Computing and Computer Assisted Interventions (MICCAI) conference, which annually hosts various medical imaging competitions that draw research teams internationally, including the BraTS challenge \cite{menze2014multimodal}. 
This task addresses a critical gap in current research—developing robust models that can generalize across domain shifts introduced by demographic, anatomical, and acquisition variability. Compared to previous years that focused on Global North adult glioma segmentation \cite{bakas2017advancing, bakas2017segmentation}, this challenge prioritizes equity in AI development by evaluating segmentation methods on data from underserved populations (e.g., in sub-Saharan Africa), where automated solutions could provide the most clinical benefit.

Deep learning remains the standard for brain tumor segmentation, with most top-performing solutions based on U-Net-style architectures since the BraTS 2014 challenge \cite{unet, ferreira2024wonbrats2023adult}. Recent work has focused on enhancing these architectures through better feature encoding, skip connections, attention mechanisms, and normalization strategies. One such advancement is the MedNeXt architecture, which adapts ConvNeXt blocks into a 3D U-Net-like framework \cite{liu2022convnet, roy2023mednext}.

In this paper, we present our state-of-the-art segmentation pipeline \textbf{\textit{EMedNeXt}} based on MedNeXt V2, designed specifically for the SSA task in BraTS-Lighthouse 2025. \textbf{\textit{EMedNeXt}} introduces three key improvements over its predecessor: (1) a larger region of interest (ROI) for better contextual learning, (2) an updated architectural skeleton inspired by nnU-Net v2, and (3) a framework for ensembling the model to enhance prediction robustness. We also explore deep supervision, training optimizations, and post-processing strategies tailored to this domain. The data used for performance assessment was obtained from standard clinical care for brain tumors and was annotated by radiologists and reviewed by neurologists to ensure accuracy \cite{africadata, bratsafrica2025}. Our results on the hidden validation set that exceeded the best model of last year demonstrate the effectiveness of our pipeline in addressing real-world distribution shifts and resource constraints. The remainder of this paper is organized as follows. The methods, including the datasets used, the architecture, and the framework training flow, are discussed in section \ref{sec:Methods}. The results and discussion, including the performance evaluation, are presented in section \ref{sec:results}. Finally, the conclusion and future work are summarized in section \ref{sec:conc}.

\section{Methods}
\label{sec:Methods}
In this section, we outline the datasets used, the architectural backbone of our model, and the comprehensive training paradigm employed to build our segmentation system.

\subsection{Data}
\label{sec:data}
In this study, we utilize two different datasets to train and evaluate our tumor segmentation pipeline, both provided as part of the BraTS-Lighthouse 2025 Challenge. Due to the limited number of samples (60 patients) in the Sub-Saharan African (SSA) dataset (Task \#5), we also incorporate data from the Pre- and Post-Treatment Adult Glioma (PPTAG) dataset (Task \#1), with a uniform size of $240{\times}240{\times}155$.

\subsubsection{2.1.1 BraTS Sub-Saharan African Dataset}
The SSA dataset consists of multi-parametric MRI scans acquired from clinical sites in sub-Saharan Africa, specifically Nigeria. Each case includes four MRI modalities—T1-weighted (T1), T1 with contrast enhancement (T1c), T2-weighted (T2), and Fluid-Attenuated Inversion Recovery (FLAIR)—along with expert-annotated tumor segmentation masks. This dataset captures the real-world variability and noise introduced by low-resource imaging settings, such as low-field MRI scanners and diverse patient demographics. It includes 60 patients' training and 35 validation brain MRI scans. An example case is shown in Figure~\ref{fig:ssa-example}.
\subsubsection{2.1.2 Pre- and Post-Treatment Adult Glioma Dataset}
\begin{figure}[!h]
    \centering
    \begin{minipage}{0.9\linewidth}
        \centering
        \includegraphics[width=\linewidth]{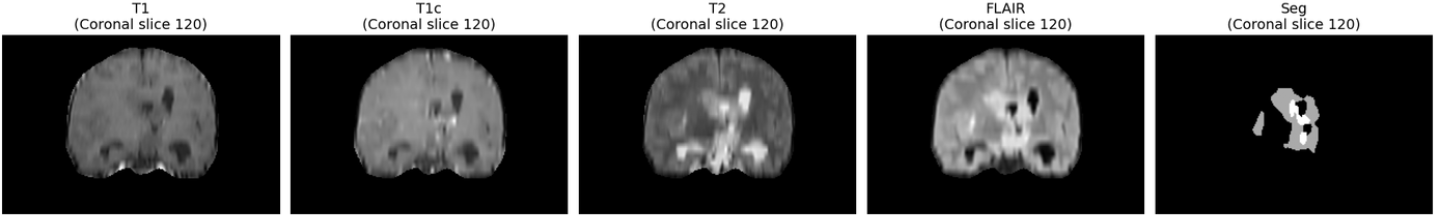}
        \caption{Cross sections of the four modalities obtained from a sample data-point from the SSA dataset along with the corresponding segmentation masks}
        \label{fig:ssa-example}
    \end{minipage}
    
    \vspace{1em} 

    \begin{minipage}{0.9\linewidth}
        \centering
        \includegraphics[width=\linewidth]{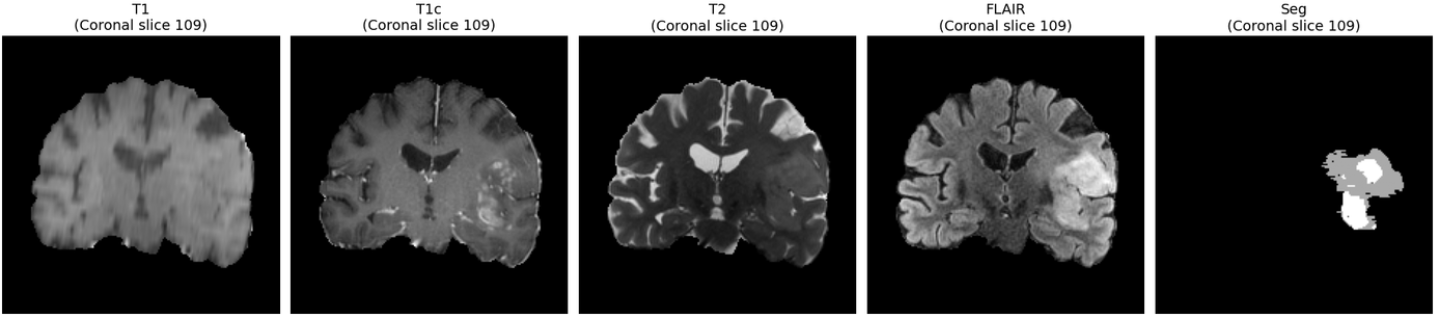}
        \caption{Cross sections of the four modalities obtained from a sample data-point from the PPTAG dataset along with the corresponding segmentation masks.}
        \label{fig:pptag-example}
    \end{minipage}
\end{figure}

The PPTAG dataset contains high-quality MRI scans of adult glioma patients captured before and after treatment. Like the SSA dataset, it includes all four standard modalities and expert annotations. However, due to differences in imaging quality, the dataset underwent a preprocessing pipeline involving denoising, intensity normalization, and spatial resampling to better align with the SSA distribution. Given the limited number of SSA cases, we augment our training set by merging the PPTAG samples with SSA data. It includes 1195 training and 219 validation brain MRI scans. An example from the PPTAG dataset is shown in Figure~\ref{fig:pptag-example}.

\subsection{MedNeXt V2}
Having established the datasets, we now describe the core segmentation architecture used in \textbf{\textit{EMedNeXt}}: MedNeXt V2, which enhances the original MedNeXt V1 for improved accuracy and robustness through several key upgrades:
\begin{figure}[!h]
    \centering
    \includegraphics[width=0.9\linewidth]{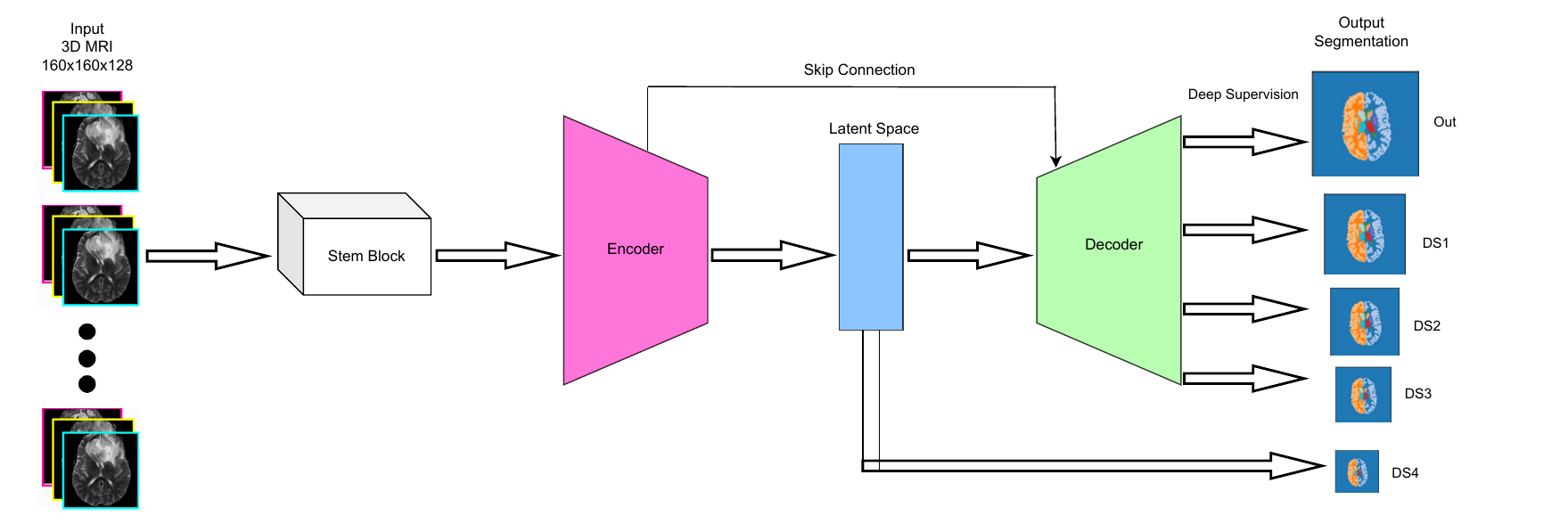}
    \caption{MedNeXt V2: High-level architecture. A 4-channel 3D MRI input ($160\times160\times128$) is processed through a stem block, followed by an encoder that extracts hierarchical features, a bottleneck, and a decoder that reconstructs segmentation maps.}

    \label{fig:mednext-fig}
\end{figure}
\begin{figure}[H]
    \centering
    \includegraphics[width=0.8\linewidth]{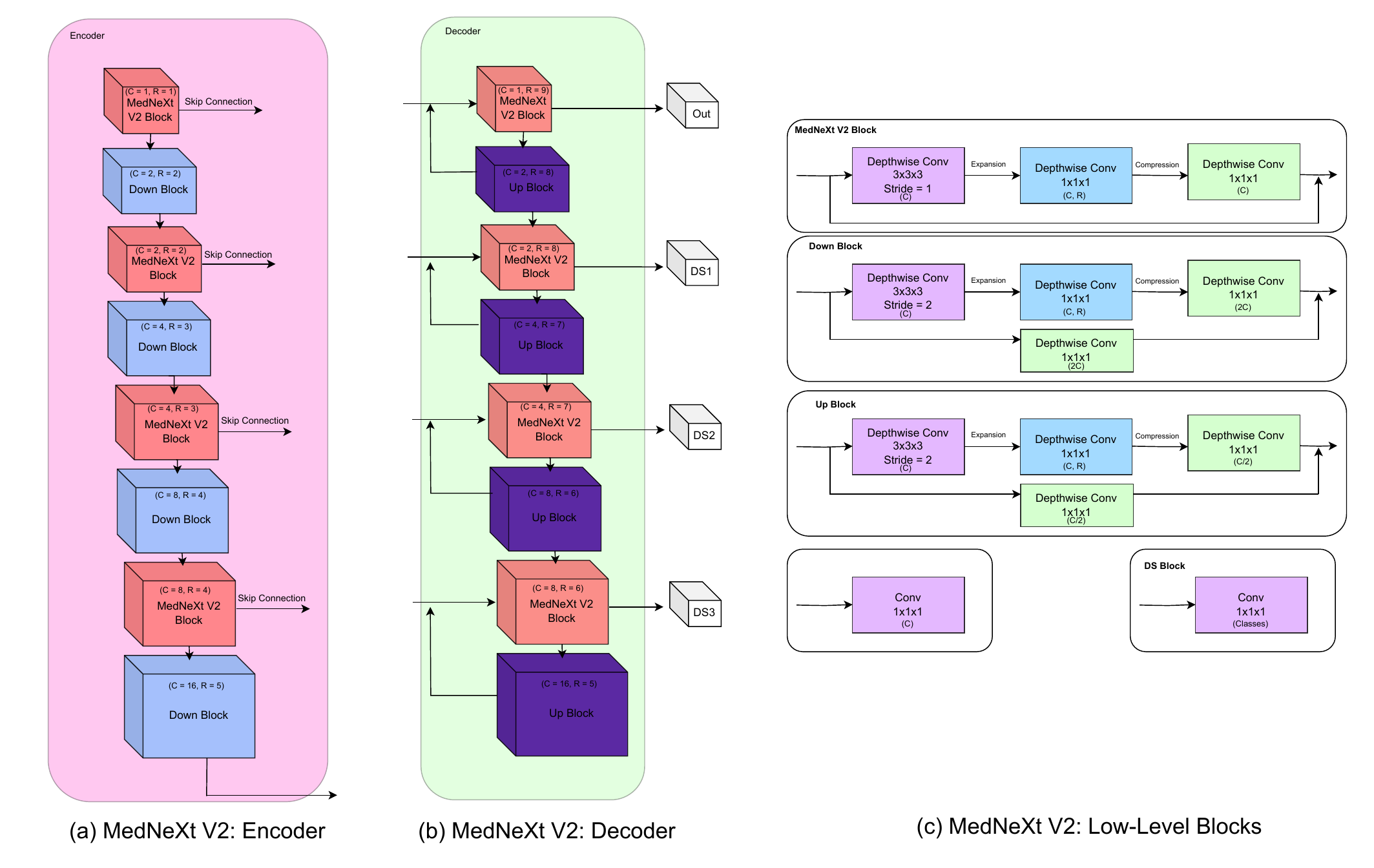}
    \caption{Detailed MedNeXt V2 architecture: (a) encoder path, (b) decoder path, and (c) low-level block components.}
    \label{fig:mednext-overall-fig}
\end{figure}

\begin{enumerate}[label=(\roman*)]
    \item \textbf{Adopting nnU-Net V2:} A dynamic, data-driven framework for segmentation without manual tuning.
    
    \item \textbf{Kernel Size Unification:} All depthwise convolutions now use a fixed $3 \times 3 \times 3$ kernel instead of variable kernel sizes with channel expansion ($C \times R$) and compression ($C$) are explicitly performed using $1 \times 1 \times 1$ convolutions after depth-wise convolutions.

    \item \textbf{Larger Field of View:} The input patch size is increased to $160 \times 160 \times 128$ to provide the network with more contextual information.
    
    \item \textbf{Deep Supervision:} Enhanced Auxiliary outputs are introduced at multiple decoder levels to improve gradient flow and training stability.

    \item \textbf{Improved Skip Connections:} Enhanced skip connections are used to better preserve spatial alignment between encoder and decoder features.
\end{enumerate}
Figure~\ref{fig:mednext-fig} provides the high-level layout; architectural details of the encoder, decoder, and block design appear in Fig.~\ref{fig:mednext-overall-fig}c. The encoder (Figure \ref{fig:mednext-overall-fig}a) is composed of alternating MedNeXt V2 blocks and Down blocks, each progressively reducing spatial resolution while increasing feature depth. The decoder (Figure \ref{fig:mednext-overall-fig}b) symmetrically mirrors the encoder using Up blocks followed by MedNeXt V2 blocks. It integrates skip connections from the encoder and produces four outputs: one final segmentation map and three intermediate outputs for deep supervision.

\subsection{EMedNeXt Training Flow}
\begin{figure}[!h]
    \centering
    \includegraphics[width=0.75\linewidth]{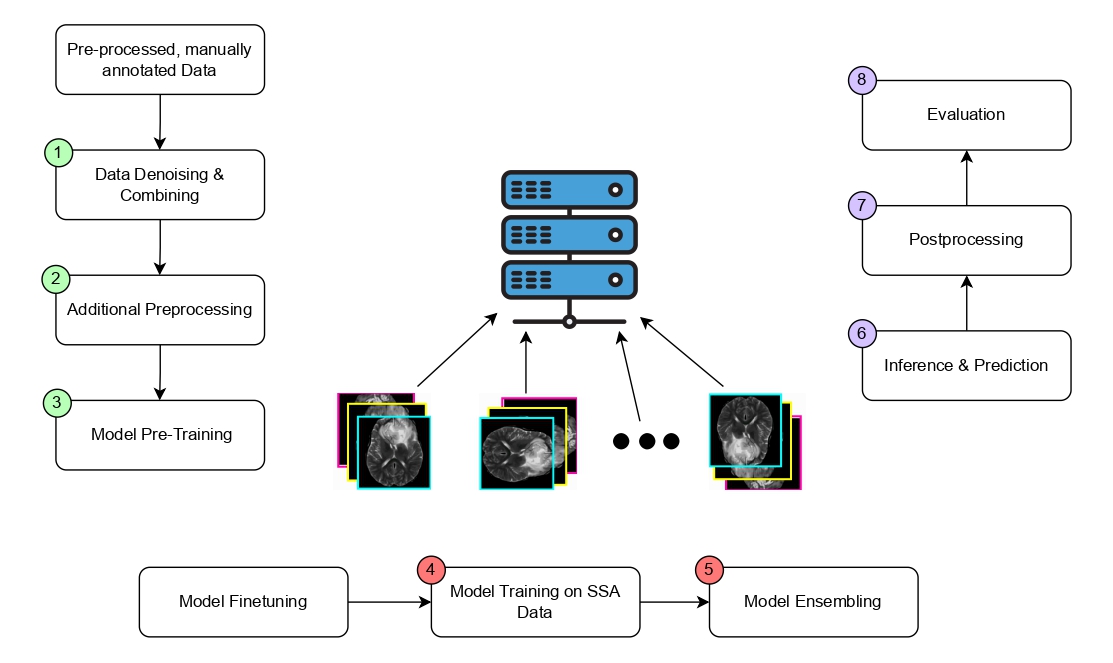}
    \caption{EMedNeXt: Training Flow} 
    \label{fig:mednext-train-fig}
\end{figure}

Having outlined the architecture, we now present our complete training workflow for \textbf{\textit{EMedNeXt}}. The pipeline is designed to effectively handle heterogeneous and limited-resource MRI datasets. Figure~\ref{fig:mednext-train-fig} illustrates the two-phase training paradigm: initial pre-training using diverse datasets followed by specialized fine-tuning on the SSA dataset. The robustness of this pipeline is further enhanced through additional steps including model ensembling, inference, postprocessing, and evaluation.

\subsubsection{2.3.1 Data Denoising \& Combining}

To prepare the training dataset, we merged the SSA dataset and the PPTAG dataset, as detailed in Subsection~\ref{sec:data}. This integration addresses the limited size of the SSA dataset by incorporating additional annotated examples while aiming to preserve domain relevance. The raw NIfTI volumes from all patients were first denoised and intensity-standardized. Each modality (FLAIR, T1, T1c, T2) was loaded and cast to \texttt{int16}, with negative outliers clipped and extreme values reset to zero to suppress scanner artifacts. To harmonize the intensity distributions, we applied channel-wise normalization limited to nonzero voxels, ensuring that each modality contributes comparably to the downstream model and that background regions do not skew the normalization statistics. Finally, all image volumes were resampled to a fixed voxel spacing of $(1.0, 1.0, 1.0)$ and a target shape of $160 \times 160 \times 128$ using cubic interpolation. The resulting unified dataset contained 1,255 preprocessed training cases with corresponding multi-class tumor segmentation masks.

\subsubsection{2.3.2 Additional Preprocessing}
Following denoising and normalization, we applied structural preprocessing to localize and standardize the region of interest. Using MONAI's spatial bounding box utilities, we identified the tightest enclosing box around all nonzero voxels in each scan and cropped both the image and its corresponding segmentation label accordingly. To ensure uniform input size for training, the cropped images were padded (or further cropped if needed) to $160 \times 160 \times 128$ using symmetric padding centered on the brain. Next, all four MRI modalities were stacked into a single 4-channel volume. We also added a fifth channel representing the aggregated foreground mask across modalities, which helps the model distinguish brain tissue from background. Segmentation masks underwent identical transformations to maintain spatial alignment. All outputs, preprocessed images, masks, and metadata (such as bounding boxes and original shapes), were stored as NumPy arrays in a designated directory. To accelerate the pipeline, we parallelized the process using Python's \texttt{multiprocessing} library across CPU cores.

\subsubsection{2.3.3 Model Pre-Training}
As we described earlier, we first merge and preprocess the SSA and PPTAG datasets to increase training diversity. Building on this, we perform a preliminary pre-training stage on the combined dataset to provide the model with a broader distribution of tumor cases and imaging conditions. This stage serves to initialize the model with robust feature representations, which led to an effective fine-tuning when the model is later trained on the SSA dataset.

\subsubsection{2.3.4 Data Augmentation}
To enhance the robustness and generalization of the model, we applied both spatial and intensity-based augmentations during training. Spatial augmentations included random cropping of regions of interest (ROIs) with dimensions $(r_x, r_y, r_z)$ and random flips along the sagittal, coronal, and axial axes with a probability of $p=0.5$, thereby increasing anatomical variability. Intensity augmentations consisted of random scaling (factor $\pm 0.1$) and shifting (offset $\pm 0.1$) of image intensities, each applied with probability $p=1.0$, to simulate inter-patient and inter-scanner variations. Collectively, these augmentations improve model generalization while preserving clinically relevant features.
\subsubsection{2.3.5 Model Finetuning \& Training on SSA Dataset}
Given the limited size and domain shift of the SSA dataset, we adopt a lightweight adaptation strategy that focuses training primarily on the \emph{decoder}. This allows us to preserve \emph{encoder} features learned during pre-training while tailoring the decoder to SSA-specific imaging artifacts and tumor patterns. Hence, we implemented a \emph{structured freezing strategy}:
\begin{enumerate}[label=(\roman*)]
    \item Freeze \emph{all} encoder parameters to retain generic representations.
    \item Unfreeze the last $k$ decoder blocks, all upsamplers, and the 3 segmentation heads (main and two deep supervision outputs).
    \item Optionally, unfreeze the deepest 2 encoder stages to enable minor low-level domain adaptation.
\end{enumerate}
This setup ensures that a small subset of the model remains trainable, typically no more than $X\% \ (X \in \{34.8 \%,38.7\%\})$ of the total parameters, helping avoid overfitting. We, then, optimize the model using AdamW Schedule-Free~\cite{schedulefree}, with distinct learning rates for different parameter groups:
\begin{itemize}
    \item Main body: $\eta = 1.0 \times 10^{-4}$
    \item Decoder blocks: $\eta_{\text{dec}} = \eta$
    \item Segmentation heads: $\eta_{\text{head}} = 2\eta$
    \item Unfrozen encoder stages (optional): $\eta_{\text{enc}} = 0.1\eta$
\end{itemize}
This configuration removes the need for explicit learning-rate schedules, as the optimizer adapts the step size automatically. 
Training proceeds in two phases: (i) \textbf{pre-training} on the merged SSA+PPTAG dataset for 150 epochs, and (ii) \textbf{fine-tuning} on the SSA dataset for 50 epochs under the structured freezing strategy. Both stages use $160{\times}160{\times}128$ voxel patches, batch size 3, and mixed precision.


To enhance boundary quality, which is crucial for \textit{Normalised Surface Dice} (NSD), we utilized a hybrid boundary‑aware loss: the standard Dice–Focal loss with a 3D boundary term. Let $\mathbf{p} \in [0,1]^{C \times H \times W \times D}$ be the predicted probabilities and $\mathbf{g} \in \{0,1\}^{C \times H \times W \times D}$ be the one-hot ground truth. The total loss becomes:
\begin{equation}
    \mathcal{L}_{\text{total}} = \mathcal{L}_{\text{Dice–Focal}} + \alpha\,\mathcal{L}_{\text{boundary}}, \qquad \alpha = 0.5
\end{equation}
The boundary loss term is defined as:
\begin{align}
    \mathcal{L}_{\text{boundary}} = \left\| \nabla_{\text{Sobel}}\mathbf{p} - \nabla_{\text{Sobel}}\mathbf{g} \right\|_2^2
\end{align}
where $\nabla_{\text{Sobel}}$ denotes a 3D Sobel operator applied channel-wise. The total loss is computed across all decoder outputs with deep supervision weights $w_i = 2^{-i}$ (finest resolution $i{=}0$), emphasizing the finer-resolution outputs.

\subsubsection{2.3.6 Model Ensembling and Inference}
Following the fine-tuning stage, we further boost segmentation robustness by aggregating predictions from multiple MedNeXt variants. As observed in preliminary experiments, individual checkpoints tend to inconsistently over- or under-segment certain tumor sub-regions, which ensembling helps mitigate. To achieve this, we apply late fusion at the probability level, where each model $m$ produces a soft probability map $P_{m} \in [0,1]^{C \times D \times H \times W}$ over the $C=3$ BraTS tumor classes: whole tumor (WT), tumor core (TC), and enhancing tumor (ET). Each model is assigned a non-negative weight vector $\mathbf{w}_{m} = (w_{m,1}, w_{m,2}, w_{m,3})$, and the ensemble probability for voxel $(x,y,z)$ and class $c$ is calculated via a weighted average:
\begin{equation}
  \hat{P}_{c}(x,y,z) = 
  \frac{\sum_{m=1}^{M} w_{m,c}\,P_{m,c}(x,y,z)}{\sum_{m=1}^{M} w_{m,c}},
  \qquad c \in \{\mathrm{WT}, \mathrm{TC}, \mathrm{ET}\}.
  \label{eq:ensemble}
\end{equation}
In our implementation, we use uniform weights $\mathbf{w}_{m} = (1,1,1)$ across all models, yielding an arithmetic mean.\footnote{The framework allows class- or model-specific weighting, useful for discounting noisy channels or low-performing models, by adjusting $w_{m,c}$ in Eq.~\eqref{eq:ensemble}.} To reduce memory requirements and support arbitrary ensemble sizes, we adopt a two-pass inference strategy:

\begin{enumerate}[label=(\roman*)]
    \item \textbf{Per-model inference.} Each checkpoint is processed independently using sliding-window inference (patch size $160{\times}160{\times}128$, with $50\%$ overlap), combined with 7-way test-time augmentation (TTA) via flipping.
    
    \item \textbf{Normalization pass.} Once all $M$ models have been evaluated, we normalize the accumulated predictions using a multiprocessing pool, dividing by the sum of weights as per Eq.~\eqref{eq:ensemble}, yielding the final ensemble output $\hat{P} \in \mathbb{R}^{3 \times D \times H \times W}$.
\end{enumerate}
This strategy requires only the memory of a single model at runtime and avoids synchronization overhead. Also, the modular design enables seamless integration of additional checkpoints in future ensemble configurations.

\subsubsection{2.3.7 Postprocessing}
To further refine the segmentation predictions obtained from the ensemble stage, we apply a tailored postprocessing pipeline that restores anatomical consistency and suppresses residual false positives. The ensemble inference outputs a three-channel probability tensor  
\[
\mathbf{P}=\bigl\{P_{\text{TC}},P_{\text{WT}},P_{\text{ET}}\bigr\} \in [0,1]^{3\times D\times H\times W},
\]
for each subject, representing soft predictions for the TC, WT, and ET classes. Although the finetuned MedNeXt V2 backbone incorporates deep supervision, raw outputs still contain numerous
sub‑millimetric false positives. The postprocessing pipeline described below mitigates these issues:
\begin{enumerate}[label=(\roman*)]
    \item \textbf{Independent hard thresholding:} Each probability map is binarized using class-specific thresholds:
    \[
    \tau_{\text{TC}} \in \{0.5, 0.7\}, \quad \tau_{\text{WT}} \in 0.5, \quad \tau_{\text{ET}} \in \{0.5, 0.7\}.
    \]
    The elevated threshold for ET reduces low-confidence activations that would otherwise degrade the NSD score.

    \item \textbf{Connected-component (CC) pruning:} For each class $c \in \{\text{TC}, \text{WT}, \text{ET}\}$, we extract 26-connected components $\mathcal{C} = \{C_1, \dots, C_K\}$ and retain components satisfying:
    \begin{equation}
    |C_k| \ge \gamma_c \quad \text{and} \quad \bar P_c(C_k) = \frac{1}{|C_k|} \sum_{x \in C_k} P_c(x) \ge \eta_c,
    \label{eq:ccFilter}
    \end{equation}
    where $(\gamma_c, \eta_c) = (150, 0.1)$ for TC, $(500, 0.1)$ for WT, and $(100, 0.1)$ for ET. If more than 10 components pass the filter~\eqref{eq:ccFilter}, we retain only the largest 10 per class.

    \item \textbf{Hierarchical enforcement:} To ensure topological correctness, we enforce the anatomical hierarchy $\text{ET} \subseteq \text{TC} \subseteq \text{WT}$ by propagating accepted ET masks into TC, and TC into WT. We then reapply the connected-component filter~\eqref{eq:ccFilter} to restore structural coherence and improve boundary-sensitive metrics.

    \item \textbf{Label fusion with priority rules:} Finally, the three refined binary masks are merged into a single label map $S \in \{0,1,2,3\}^{D \times H \times W}$ using a fixed priority: ET~$\triangleright$~TC~$\triangleright$~WT:
    \[
    S(x) =
    \begin{cases}
    3 & \text{if } \text{ET}(x) = 1, \\
    2 & \text{else if } \text{TC}(x) = 1, \\
    1 & \text{else if } \text{WT}(x) = 1, \\
    0 & \text{otherwise}.
    \end{cases}
    \]
\end{enumerate}

\subsubsection{2.3.8 Evaluation \& Experimental Setup}
To ensure fair evaluation, we followed the official BraTS 2025 SSA protocol. During model development, we performed $K$-fold cross-validation ($K{=}5$) on the SSA training set (60 cases), using 80\% of cases for training and 20\% for internal validation in each fold. This procedure helped stabilize training and guide hyperparameter selection. For final reporting, models pre-trained on the combined SSA+PPTAG training data were retrained on the full SSA training set and evaluated on the hidden BraTS-SSA validation set (35 cases). We assess the segmentation quality using the DSC and NSD metrics, both at global and lesion-wise levels, as defined in the challenge guidelines. All training and inference experiments were conducted on a cluster with 4 NVIDIA A6000 GPUs using mixed precision. We trained our model using the AdamW optimizer with a learning rate of $2.8e^{-3}$ and ScheduleFree scheduling for a total of 150 epochs of pre-training followed by 50 epochs of fine-tuning, enabling deep supervision. Other key configurations include a batch size of $3$, weight decay of $1e^{-6}$, and using MedNeXt-B variant with a kernel size of $3$.

\section{Results \& Discussion}
\label{sec:results}
\begin{figure}
    \centering
\includegraphics[width=0.5\linewidth]{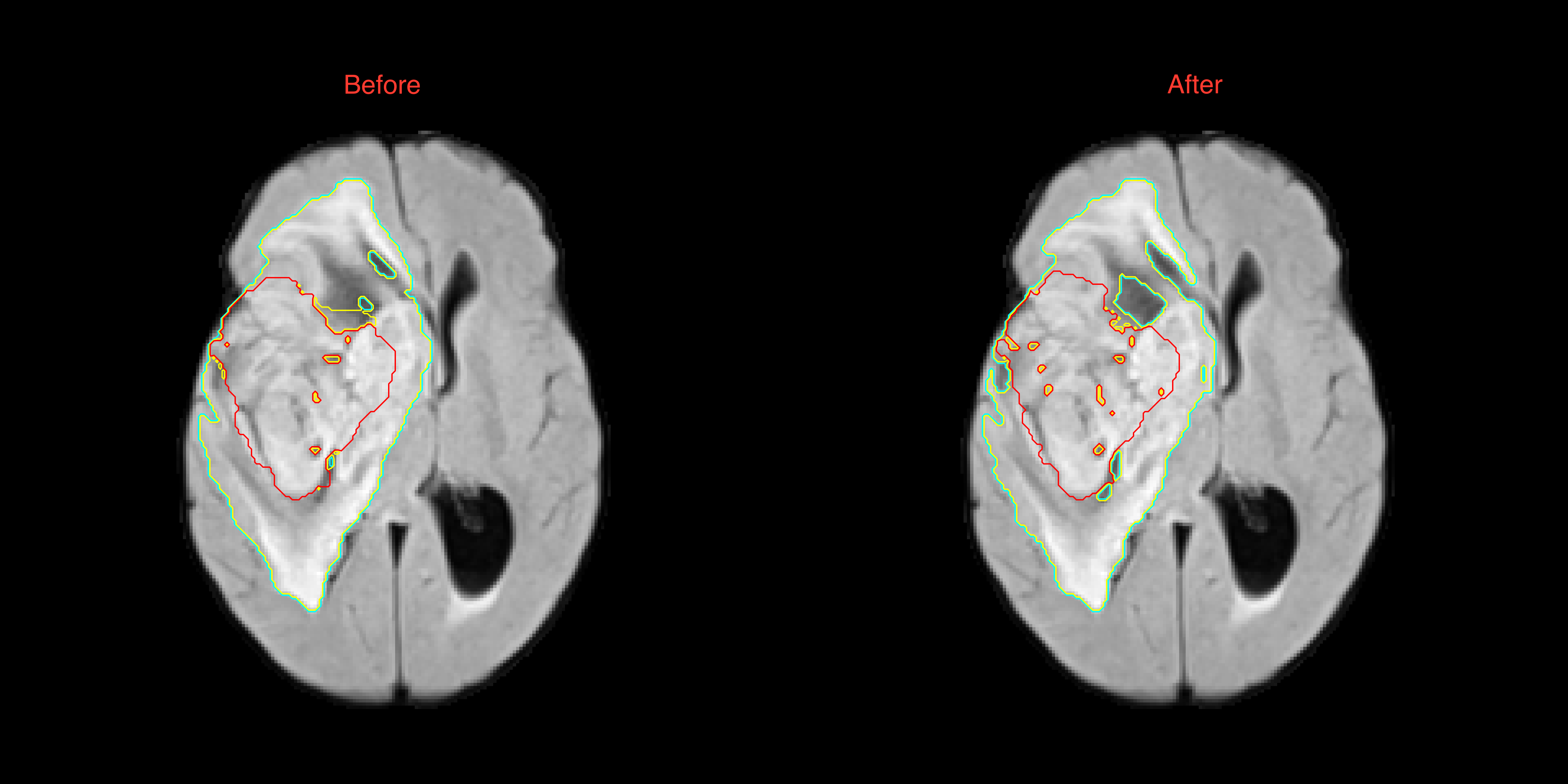}
    \caption{Comparison between Finetuned MedNext with B=5 and our final ensemble of models qualitative predictions}
    \label{fig:qualitative-results}
\end{figure}

With the training setup and evaluation methodology established, we now present the results of our experiments. We analyze the effectiveness of each component in our pipeline, from architectural upgrades to post-processing refinements.
\vspace{-0.3cm}
\begin{table}[h]
  \centering
  \caption{Lesion‑wise segmentation performance on the BraTS‑SSA hidden validation set and our best performing model on the test set (higher is better).  Best values are \textbf{bolded}.
}
  \label{tab:performance_comparison_lesionwise}
  \setlength{\tabcolsep}{3pt}
  \sisetup{
    detect-weight  = true,
    detect-inline-weight  = math,
    table-format  = 1.3, 
    round-mode= places,
    round-precision = 3
  }
  \begin{threeparttable}
  \begin{tabular}{
      @{}l
      S S S 
      S S S 
      S S S  
      @{}
  }
    \toprule
    \multirow{2}{*}{\textbf{Model}} &
      \multicolumn{3}{c}{\textbf{Dice}\,$\uparrow$} &
      \multicolumn{3}{c}{\textbf{NSD\textsubscript{0.5}}\,$\uparrow$} &
      \multicolumn{3}{c}{\textbf{NSD\textsubscript{1.0}}\,$\uparrow$} \\
    \cmidrule(lr){2-4}\cmidrule(lr){5-7}\cmidrule(lr){8-10}
        & {ET} & {TC} & {WT}
        & {ET} & {TC} & {WT}
        & {ET} & {TC} & {WT} \\
    \midrule
    Baseline (MedNeXt V1)            & 0.822 & 0.815 & 0.881 & 0.424 & 0.378 & 0.383 & 0.764 & 0.698 & 0.728 \\
    MedNeXt V2 (B{=}5)               & 0.845 & 0.860 & 0.914 & 0.501 & 0.470 & 0.444 & 0.822 & 0.776 & 0.796 \\
    FT MedNeXt V2 (B{=}5)            & 0.870 & 0.863 & 0.919 & 0.570 & 0.513 & 0.499 & 0.863 & 0.798 & 0.821 \\
    MedNeXt V2 (B{=}3)               & 0.860 & 0.832 & 0.904 & 0.569 & 0.479 & 0.487 & 0.852 & 0.763 & 0.809 \\
    FT MedNeXt V2 (B{=}3)            & 0.873 & 0.835 & 0.927 & 0.569 & 0.472 & 0.498 & 0.869 & 0.769 & 0.830 \\
    Ensemble (B{=}3)                 & 0.883 & 0.873 & 0.933 & 0.579 & 0.519 & 0.520 & 0.873 & 0.806 & 0.839 \\
    \textbf{Ensemble + PP}\tnote{*}  & \bfseries 0.883 & \bfseries 0.873 & \bfseries 0.933
                                     & \bfseries 0.580 & \bfseries 0.522 & \bfseries 0.521
                                     & \bfseries 0.873 & \bfseries 0.806 & \bfseries 0.839 \\
    \midrule
    \multicolumn{10}{c}{\textbf{Test-phase Results}} \\
    \rowcolor{tmi_blue!10} \textbf{Ensemble + PP (Mean)} & 
0.8597 & 0.8821 & 0.9298 & \text{-} & \text{-} & \text{-} & 0.8534 & 0.8182 & 0.8544 \\
\rowcolor{tmi_red!10} \textbf{Ensemble + PP (Std)} & 
0.1612 & 0.1749 & 0.1147 & \text{-} & \text{-} & \text{-} &  0.1558 & 0.1908 & 0.1316 \\
    \bottomrule
  \end{tabular}
  \begin{tablenotes}[flushleft]
    \footnotesize
    \item[*] Post‑processing: thresholds $\tau_{\text{ET}}=\tau_{\text{TC}}=0.625$ and
             connected‑component filter $|CC_{\text{ET}}|\ge30$ voxels.
  \end{tablenotes}
  \end{threeparttable}
\end{table}
Table~\ref{tab:performance_comparison_lesionwise} summarizes the lesion‑wise Dice Similarity Coefficient (DSC) and Normalized Surface Dice (NSD) achieved by successive versions of our framework.  Starting from the original \textit{MedNeXt V1} baseline, each component we introduced gives a boost in our segmentation quality:

\begin{enumerate}[label=(\roman*)]
    \item \textbf{Backbone upgrade (V1 $\rightarrow$ V2).}  
          Replacing the V1 encoder–decoder with the larger‐receptive‑field \textit{MedNeXt V2} architecture (``Base= 5'') lifts the mean DSC from \textbf{0.839} to \textbf{0.873} and the mean NSD$_{0.5}$ from \textbf{0.395} to \textbf{0.472}. The gains are maximized for enhancing‑tumour (ET) class.

    \item \textbf{Domain‑adaptive fine‑tuning.}  
          Freezing the encoder and re‑training only the decoder on the SSA dataset with the boundary loss improves boundary alignment with the ground-truth. Mean DSC is slightly pushed to \textbf{0.884}, and mean NSD$_{0.5}$ to \textbf{0.518}. Therefore, fine‑tuning allows the model to learn SSA‑specific intensity profiles while avoiding over‑fitting to the small training set.

    \item \textbf{Base architecture variant with kernel size of 3.}  
          Narrowing the channel width gave us a boost in performance. After fine‑tuning, the ``Base=3'' model attains a DSC of \textbf{0.878} and an NSD$_{0.5}$ of \textbf{0.513}.

    \item \textbf{Ensembling.}  
          Averaging the logits of 5 our best base with fine-tuned checkpoints of the kernel size 3 model pushed the mean DSC to \textbf{0.896} and the NSD$_{0.5}$ to \textbf{0.537}. Ensemble voting mitigates individual checkpoint biases.

    \item \textbf{Post‑processing optimization.}  
       Adjusting the hard‑threshold for ET/TC to $0.625$ and relaxing the ET component‑size filter (from 100 to 30 voxels) recovers additional small lesions, raising the final NSD$_{0.5}$ to \textbf{0.541} and NSD$_{1.0}$ to \textbf{0.84}, while leaving the already high DSC unchanged. Similarly, we used these settings along with our ensembling approach for our final submission in the test-phase.
\end{enumerate}
In general, the proposed \textbf{\textit{EMedNeXt}} pipeline delivers an average Lesion-Wise DSC of \textbf{0.897} with solid boundary alignment performance. These gains translate to visibly cleaner segmentation and fewer noisy fragments, as illustrated in Fig.~\ref{fig:qualitative-results}.
The qualitative sample highlights the advantages of the ensemble of models (right) over the single fine‑tuned \textit{B5} model (left).  The \textit{B5} model fails to capture several small, isolated enhancing tumor regions (orange contours) along the inferior margin of the lesion and within the necrotic core. These regions are consistently recovered by the ensemble, reflecting the improvement in ET-wise NSD reported in Table~\ref{tab:performance_comparison_lesionwise}.
Furthermore, the set provides a more precise whole tumor envelope (green), with noticeably tighter adherence to the actual tumor boundaries.

\section{Conclusion}
\label{sec:conc}

In this work, we presented \textbf{\textit{EMedNeXt}}, a MedNeXt V2-based framework for brain tumor segmentation tailored to the BraTS-Lighthouse 2025 SSA task. Our pipeline integrates a larger field of view, structured decoder fine-tuning via encoder freezing, and a class-specific post-processing strategy. To address domain shifts and data scarcity, we employed pretraining on PPTAG data followed by fine-tuning on SSA scans. Ensembling and optimized post-processing further improved lesion boundary quality. Our final model achieved an average LesionWise Dice of \textbf{0.897}, with NSD scores of \textbf{0.541} (0.5mm) and \textbf{0.84} (1.0mm), demonstrating strong performance in low-resource settings. Future work will extend this framework to pediatric data and explore new architectures for the segmentation framework.



\newpage
\bibliographystyle{splncs04}

\begin{thebibliography}{99}

\bibitem{gliomas}
Mesfin, F. B. and others: Gliomas. \textit{StatPearls [Internet]}, StatPearls Publishing, Treasure Island (FL), 2024. Updated 2023 May 20. [Online]. Available: \url{https://www.ncbi.nlm.nih.gov/books/NBK441874/}

\bibitem{ferreira2024wonbrats2023adult}
Ferreira, A., Solak, N., Li, J., Dammann, P., Kleesiek, J., Alves, V., Egger, J.:
How we won BraTS 2023 Adult Glioma challenge? Just faking it! Enhanced Synthetic Data Augmentation and Model Ensemble for brain tumour segmentation, 2024. [Online]. Available: \url{https://arxiv.org/abs/2402.17317}

\bibitem{unet}
Ronneberger, O., Fischer, P., Brox, T.:
U-Net: Convolutional Networks for Biomedical Image Segmentation. In: Navab, N., Hornegger, J., Wells, W. M., Frangi, A. F. (eds.), \textit{Medical Image Computing and Computer-Assisted Intervention -- MICCAI 2015}, Springer, Cham, pp. 234--241, 2015.

\bibitem{menze2014multimodal}
Menze, B. H., Jakab, A., Bauer, S., Kalpathy-Cramer, J., Farahani, K., Kirby, J., Burren, Y., Porz, N., Slotboom, J., Wiest, R., et al.:
The multimodal brain tumor image segmentation benchmark (BRATS). \textit{IEEE Transactions on Medical Imaging}, vol. 34, no. 10, pp. 1993--2024, 2014.

\bibitem{africadata}
Adewole, M., Rudie, J. D., Gbadamosi, A., Toyobo, O., Raymond, C., Zhang, D., Omidiji, O., Akinola, R., Suwaid, M. A., Emegoakor, A., et al.:
The Brain Tumor Segmentation (BraTS) Challenge 2023: Glioma Segmentation in Sub-Saharan Africa Patient Population (BraTS-Africa), 2023. [Online]. Available: \url{https://arxiv.org/abs/2305.19369}

\bibitem{schedulefree}
Defazio, A., Mehta, H., Mishchenko, K., Khaled, A., Cutkosky, A., et al.:
The Road Less Scheduled. \textit{arXiv preprint arXiv:2405.15682}, 2024.

\bibitem{owrangi2018mri}
Owrangi, A. M., Greer, P. B., Glide-Hurst, C. K.:
MRI-only treatment planning: benefits and challenges. \textit{Physics in Medicine \& Biology}, vol. 63, no. 5, p. 05TR01, 2018.

\bibitem{bakas2017advancing}
Bakas, S., Akbari, H., Sotiras, A., Bilello, M., Rozycki, M., Kirby, J. S., Freymann, J. B., Farahani, K., Davatzikos, C.:
Advancing the cancer genome atlas glioma MRI collections with expert segmentation labels and radiomic features. \textit{Scientific Data}, vol. 4, no. 1, pp. 1--13, 2017.

\bibitem{bakas2017segmentation}
Bakas, S., Akbari, H., Sotiras, A., Bilello, M., Rozycki, M., Kirby, J., Freymann, J., Farahani, K., Davatzikos, C.:
Segmentation labels and radiomic features for the pre-operative scans of the TCGA-LGG collection. \textit{The Cancer Imaging Archive}, vol. 286, 2017.

\bibitem{bratsafrica}
Adewole, M., Rudie, J. D., Gbadamosi, A., Toyobo, O., Raymond, C., Zhang, D., Omidiji, O., Akinola, R., Suwaid, M. A., Emegoakor, A., et al.:
The brain tumor segmentation (BraTS) challenge 2023: Glioma segmentation in Sub-Saharan Africa patient population (BraTS-Africa). \textit{ArXiv}, 2023.

\bibitem{roy2023mednext}
Roy, S., Koehler, G., Ulrich, C., Baumgartner, M., Petersen, J., Isensee, F., Jaeger, P. F., Maier-Hein, K.:
MedNeXt: Transformer-driven Scaling of ConvNets for Medical Image Segmentation. \textit{arXiv preprint arXiv:2303.09975}, 2023.

\bibitem{liu2022convnet}
Liu, Z., Mao, H., Wu, C.-Y., Feichtenhofer, C., Darrell, T., Xie, S.:
A ConvNet for the 2020s. In: \textit{CVPR}, pp. 11976--11986, 2022.

\bibitem{bratsafrica2025}
Adewole, M., Rudie, J. D., Gbadamosi, A., et al.:
The BraTS-Africa Dataset: Expanding the Brain Tumor Segmentation (BraTS) Data to Capture African Populations. \textit{Radiology: Artificial Intelligence}, 2025. DOI: \url{10.1148/ryai.240528}

\end{thebibliography}

\end{document}